# A Material-based Panspermia Hypothesis: The Potential of Polymer Gels and Membraneless Droplets

Mahendran Sithamparam[a], Nirmell Satthiyasilan[a], Chen Chen[b], Tony Z Jia*[b,c] and Kuhan Chandru*[a]

[a]Space Science Center (ANGKASA), Institute of Climate Change, Level 3, Research Complex, National University of Malaysia (UKM), UKM Bangi, Selangor 43600, Malaysia
[b]Earth-Life Science Institute, Tokyo Institute of Technology, Meguro-ku, Tokyo 152-8550, Japan
[c]Blue Marble Space Institute of Science, Seattle, Washington 98154, United States

**Corresponding Authors:** tzjia@elsi.jp and kuhan@ukm.edu.my

**Abstract:**

The Panspermia hypothesis posits that either life's building blocks (molecular Panspermia) or life itself (organism-based Panspermia) may have been interplanetary transferred to facilitate the Origins of Life (OoL) on a given planet, complementing several current OoL frameworks. Although many spaceflight experiments were performed in the past to test for potential terrestrial organisms as Panspermia seeds, it is uncertain whether such organisms will likely "seed" a new planet even if they are able to survive spaceflight. Therefore, rather than using organisms, using abiotic chemicals as seeds has been proposed as part of the molecular Panspermia hypothesis. Here, as an extension of this hypothesis, we introduce and review the plausibility of a polymeric material-based Panspermia seed (M-BPS) theoretical concept, where the type of polymeric material that can function as a M-BPS must be able to: 1) survive spaceflight, and 2) "function", *i.e.*, contingently drive chemical evolution towards some form of abiogenesis once arriving on a foreign planet.

We use polymeric gels as a model example of a potential M-BPS. Polymeric gels that can be prebiotically synthesized on one planet (such as polyester gels) could be transferred to another planet *via* meteoritic transfer, where upon landing on a liquid bearing planet, can assemble into structures containing cellular-like characteristics and functionalities. Such features presupposed that these gels can assemble into compartments through phase separation to accomplish relevant functions such as encapsulation of primitive metabolic, genetic and catalytic materials, exchange of these materials, motion, coalescence, and evolution. All of these functions can result in the gels' capability to alter local geochemical niches on other planets, thereby allowing chemical evolution to lead to OoL events.

## 1. Introduction:

The Origins of Life (OoL) is an unsolved scientific conundrum, and there are two general ways to think about it: 1) that life started exclusively on Earth by the means of geologically driven prebiotic chemistry, *i.e.*, abiogenesis[1], or 2) that life, or some of its components (spores, building blocks of biochemicals, and inorganic chemicals or its minerals, *etc.*)[2], may have been transported to Earth (or other planets) by interplanetary[3], interstellar[4] or even intergalactic[5] transfer (*e.g.*, by meteorites, comets, stardusts *etc.*)[6–8], *i.e.*, the Panspermia hypothesis.

In Greek, the word *Pan* means "universal", and *Spermia* means "life", and were both interchangeably used as early as the 5th-century by the philosopher Anaxagoras[9]. In modern times, these words gained scientific attention when Svante Arrhenius suggested that extraterrestrial spores propelled by solar radiation may have been transferred/seeded to early Earth, *i.e.*, "radiopanspermia"[10]. Since then, many space-flight experiments (*e.g.*,[11–16]) have been conducted to test the (survivability) limits of terrestrial life during space travel, in hopes to discern the plausibility of transporting extant life to other planets (a key aspect to the Panspermia hypothesis). Though it is beyond the scope of this paper to argue how life (or its constituents, *e.g.*, biomolecules) can be transferred throughout the universe[17–19], several theoretical models have shown that Panspermia is a likely plausible phenomenon[20,21] between Earth and Mars. These theoretical estimates are mainly supported by the discovery of Martian meteorites on Earth[22], and that these meteorites could possibly endure tolerable impact shocks thereby preserving their content (*e.g.*, bacterial spores and lichens)[23]. This idea of meteorites transporting life and biomolecules to another planet is better known as "lithopanspermia".

There are many different versions of Panspermia theories (see Table 1 for summary), and they always involve the transfer of organism(s) and/or its crucial building blocks such as biochemicals (*e.g.*,[24,25]) and/or their precursors (*e.g.*,[26,27]), and inorganic compounds or minerals (*e.g.*,[25,28,29]) enclosed within meteorites or comets that were transported to (recipient) planets to start, enable and facilitate the OoL processes.

A similar version but with a touch of "intelligence", called, "directed Panspermia"[30], suggests the *deliberate* seeding of life on the Earth by an assumed hypothetical technological advanced extraterrestrial civilization[31,32]. When this narrative is reversed, we can assume that humans are *the* extraterrestrial civilization capable of performing such directed Panspermia feat, *i.e.*, "human-directed Panspermia" also popularly known as "terraforming"[33]. In fact, attempts on terraforming Mars[34,35] are making news as a possible avenue to mitigate global problems (*e.g.*, overpopulation and climate change) and by preserving the continuity of terrestrial life at the same time. A more recent but different sub-iteration of this version is "Protospermia"[36]. This version suggests that human-directed transfer of chemicals can nudge (or push) a barren planet towards abiogenesis as how contingently the process (may) see fit, but not necessarily to make life similar to that on Earth.

Although the Panspermia hypothesis has been extensively studied and reviewed in the past through all of these different points of view (*e.g.*,[15,16,37–47]), the hypotheses do not center around the necessity of primitive polymers and biopolymer existence and function on early Earth. As such, here, we incorporate a new aspect to the hypothesis by introducing the material-based Panspermia hypothesis, *i.e.*, a scenario where Panspermia seeds are made of materials (*e.g.*, polymers, gels, *etc*), and not by microbes or monomeric chemicals (*e.g.*, amino acids, nucleotides, *etc*). These prebiotic polymeric gels, we hypothesize, can be plausibly prebiotically formed on a hypothetical donor planet, may endure spaceflight, and subsequently, land on a recipient habitable planet with the ability to nudge chemical evolution in its new environment, thus prompting an OoL event (Figure 1). Hence, in this piece, we will: 1) briefly account for some notable Panspermia experiments and describe their limitations or any knowledge gaps when explaining the mechanism of OoL. And then 2) speculate about the potential of a material-based Panspermia model, in the form of prebiotic polymeric gels, that may shed some light on plausible OoL mechanisms. Such a materials-based Panspermia model could have occurred on early Earth, and has the plausibility, after further engineering and development, to become an experimental apparatus for terraforming planets in the future.

Table 1: Brief descriptions of the common types of Panspermia models

| Types of Panspermia model | Descriptions | References |
|---|---|---|
| Radiopanspermia | Microorganisms/spores, as proposed by SvanteArrhenius in 1908, escaped their donor planets that are drifting in space, carried by the force and pressure of solar radiation to interstellar space that would eventually *seed* another planet. | 48–51 |
| Lithopanspermia | Life, biomolecules and its precursors *within* comets and meteorites, by which they are protected against space degradation, that travels between planets and galaxies. | 13,52,53 |
| Accidental Panspermia | Accidental Panspermia might occur if terrestrial bacteria were to stow away on a spacecraft travelling to another planet. Hence, to avoid accidental Panspermia to other planets with potential life, such as Mars, NASA and other space institutes follow strict decontamination protocols on all interplanetary vehicles. | 54–56 |
| Directed Panspermia | Directed Panspermia means that an advanced civilization purposely and directly *seeded* life on the target planetary body. | 30,33,57–61 |
| Molecular Panspermia | The building blocks of life came from space. Simple molecules (*e.g.*, glycine) or even elements (P) could have prompted the OoL. | 3,19,62 |
| Material-based Panspermia | An extension of the molecular Panspermia that involves a more higher-ordered group of chemicals than single units of chemicals (monomers), *i.e.*, polymers, ceramics, gels and composites, that could have enabled the OoL. | This paper |

## 2. How current Panspermia theories could and could not have led to the origins of life?

From a conceptual point of view, there are generally two ways for Panspermia seeds to enable an OoL event on "habitable" planets ([63–65]), *i.e.*, the close proxies of Earth in size, environment, and distance to a star that is presumably capable of harboring life (although the term "habitable zone" has been coined for the orbital location of such planets, the correctness of such a term is still under debate[66–68]). Here, we select Earth as an example to discuss the plausibility of these two methods of Panspermia, as Earth is the only "habitable" planet that we have direct evidence from.

First, Panspermia seeds transported in the form of biological organisms (*e.g.*, radiopanspermia, lithopanspermia that carries organisms within meteorites, and directed Panspermia), could result in the development of life on another planet. To date, selected organisms in the form of dried microbes (*e.g.*, *Deinococcus spp.* spores[13,16]), fungi[69], and microscopic animals[70] have been investigated for their potential as Panspermia seeds. For example, these organisms were exposed to either ground-based spaceflight simulation and/or low-Earth orbit (LEO) to evaluate their survivability in such conditions[13], where some of the organisms showed remarkable survivability. Organisms that survive spaceflight are thus speculated to eventually land on a new "habitable" planet and begin colonization by utilizing existing organic and/or inorganic chemicals (on the new planet) to live, adapt to the new planet's geochemistry, and proliferate. This could begin, for example, with the organism's metabolic demands, *i.e.*, to remain alive and maintain themselves. In this case, such organisms' metabolic processes must rely on different types of geochemical redox related energy sources on the new planet that may be different from their accustomed electron donors (*e.g.*, $H_2$ used typically for methanogens/acetogens or organics for most eukaryotes on Earth) and acceptors (*e.g.*, $CO_2$ for methanogens/acetogens or $O_2$ for most eukaryotes on Earth)[71,72], or must manage to function on lower concentrations/fugacity of their accustomed electron acceptors/donors found on said new planet.

However, organisms utilizing available geochemical redox related energy sources on other planets may be improbable, as we must keep in mind that life-as-we-know-it intertwines several biological subsystems (metabolism, replication, cellularization, *etc.*) that affect each other in any given organism, resulting to survivability and also to innovate (for new functionalities) against the backdrop of their geochemical environment. More importantly, these subsystems have undergone the rigors of Darwinian selection, *i.e.*, evolutionary optimization, *but only against their own* geochemical environment. For example, assume that a directed Panspermia experiment to terraform an ideal habitable planet using the example of dried *Deinococcus spp.*[73] mentioned above — even if these dried microbes can survive spaceflight, once it reaches a new habitable planet, their biological subsystems will likely not function properly. This is because the conditions on the foreign recipient planet, though may be hypothetically similar to Earth, would have significant differences and would be detrimental to the survival and propagation of *Deinococcus spp.* (*e.g.*, differences in atmospheric pressure and composition, oceanic pH, temperature, *etc.*). Even present-day terrestrial life occupies specific niches (*e.g.*, Archaea prefers anoxic conditions, halophiles prefer high salinity, thermophiles live in high-temperature environments, *etc.*). Such organisms cannot generally be transferred from one terrestrial niche to another while being expected to survive and propagate. In other words, we also cannot simply assume that life-as-we-know-it being Earth-compatible to be compatible in extraterrestrial environments[74]. Hence, in a general context, it is unlikely that an organism, terrestrial or not, is capable of "seeding" a new planet, as they are already highly evolved and perhaps would need to rapidly "de-evolve" and "re-evolve" according to the new geochemical conditions to survive on the recipient planet.

On the other hand, if the conditions of the new planet are similar enough to that of Earth, then, perhaps survival and rapid evolution could happen considering that microbes can adapt[75–77] *via* biochemical or genomic adaptation. Biochemical adaptation refers to the quick induction and expression of existing genes within the organism to change its metabolism and express necessary genes to adapt with an environmental change. Examples of this is shown by mesophiles' ability to adapt to sudden temperature changes by either expressing heat-[78] or cold-shock proteins[79]. Whereas, genomic-level adaptation is more long term where genetic mutations and recombinations[80], or horizontal gene transfer[81] can make subsequent proteomic changes to adapt to new environmental changes. This sort of adaptation could, in principle, lead to speciation too (*e.g.*,[82]). Nevertheless, examples of microbial adaptations are demonstrated in food science (*e.g.*,[75]), biomedicine (*e.g.*,[83]), and environmental science (*e.g.*, [84,85]). However, a detailed discussion on whether microbes can or cannot adapt to a new planetary condition is limited, and beyond the scope of this paper.

In addition, there are other issues with utilizing organism-based Panspermia seed to explain the OoL as well. Typically, the time scale of an organism-based Panspermia seed to travel to a new planet may be too long, or would require a short travel window, for its internal biological components (such as proteins and DNA) to survive. For instance, known Martian meteorites took ~600,000 to 14 million years[86] to reach Earth, while bacterial DNA collected from ancient Antarctic ice (100,000 to 8 million years old) had a half-life of ~1.1 million years[87]. As such, assuming that degradation conditions on Earth are identical to in space (which are clearly not identical, as we can see below), if it took 14 million years (upper bound time) for such a terrestrial bacteria to travel between Earth and Mars, then roughly only 0.015% of its original DNA would be present upon arrival; however, a shorter travel time (~600,000 years) would allow ~68% of the original DNA to remain intact. Hence the travel durations play a critical role in the survivability of such biological molecules. Travel times to farther planets, such as exoplanets, would take even longer, and as such the amount of DNA and other biological molecules remaining after arrival would essentially have made such a journey unsurvivable for terrestrial organisms. It is worth noting that ancient terrestrial bacterial DNA samples in Antarctic ice[87] were subjected to various terrestrial degradation types (*e.g.*, hydrolysis, oxidation, etc.)[88] that are absent in space environments. However, spaceflight environments entail other forms of degradation factors not present on Earth (*e.g.*, UV radiation, microgravity, *etc.*) that could have been destructive to biological molecules. For example, radiation (*e.g.*, HZE ions and UV) is known to be destructive to DNA[89,90] and proteins[91,92]. Hence, because of the likely degradation of biochemical components within extant terrestrial organisms, organism-based Panspermia seeds would at best be the provider of molecular building blocks to a new planet, suggesting that the organism-based Panspermia model (to terraform other planets) would be an unnecessarily complex and uncontrollable (due to the unpredictability of the degradation of the biomolecular components of the seed organisms).

Second, the molecular Panspermia hypothesis supposes that molecular building blocks generated by abiotic sources (which themselves could eventually lead to assembly and synthesis of biopolymers (*e.g.*,[93]) can be provided *en masse* to any planet to enable the OoL, even if such building blocks were not necessarily present on the recipient planet. For example, molecular Panspermia would be one solution to "the phosphate problem": phosphates are present in central components utilized in biological replication (*e.g.*, nucleic acids) and metabolism (*e.g.*, ATP and Coenzyme A, *etc.*) but were not readily available on early Earth for the OoL[94,95] due to their insolubility in mineral form[29,96] (*e.g.*, apatite ($Ca_5(PO_4)_3(OH,F,Cl)$), hydroxylapatite ($Ca_5(PO_4)_3(OH)$), *etc.*)[97]. As such, their low prebiotic availability also renders chemical evolution and metabolism on early Earth improbable, if not impossible. However, soluble mineral forms of phosphate that were scarce in early Earth's mineralogy but do exist in reasonable abundance in meteoritic samples (*e.g.*, schreibersite ((Fe,Ni)$_3$P))[98] and on Mars (*e.g.*, chlorapatite)[29], this suggests a potential example of molecular Panspermia that may have happened in the past; *i.e.*, that such phosphate sources were delivered to Earth from extraterrestrial origin. This, however, would not discount other unknown reservoirs of phosphate that may have existed on early Earth (*e.g.*,[99–101]). Nevertheless, a molecular Panspermia event is plausible since some building blocks (*e.g.*, amino acids and sugars)[26,102–104], organic chemicals[105] and other inorganics[106–108] could survive spaceflight (intact) to the recipient planet of interest; *i.e.*, that Earth life originated from extraterrestrial chemicals, while extant Earth chemicals could lead to the development of life on other planets as well.

Hence, we suggest a polymer material-based testable molecular Panspermia model that may be more plausible than using biological seed organisms by assuming that Panspermia seeds need not be an organism but can be made of materials, *i.e.*, a material-based seed, that may be capable of not only facilitating or driving prebiotic chemical evolution on a wider range of habitable planets (meaning the planet does not have to be an equal contemporary of the Earth). These materials can be assembled completely from prebiotic chemistry (used to describe the OoL on Earth) or even man-made materials (used to explore its utility for terraforming planets). Essentially, this material-based seed, after being ejected from a (donor) planet into and through space *via* meteorites (or man-made space vessels), can fulfill two basic criteria: (1) be robust and avoid degradation while enduring space exposure, and (2) once arriving in an environment of a potential (recipient) habitable planet, the "seed" can trigger chemical activities (*e.g.*, releasing chemicals, receiving and/or interacting with extant chemicals or geochemistries, *etc.*) in its new environment. In other words, the seed blends and interacts with its new environment, thus ensuring chemical continuity towards some sort of nascent biology. A seed capable of accomplishing both basic criteria gives us a framework to explore and expand the significance and plausibility of the Panspermia hypothesis in the context of the OoL (Figure 1). As such, here, we focus on prebiotic material-based Panspermia seeds (M-BPS) as a mechanism leading to the OoL.

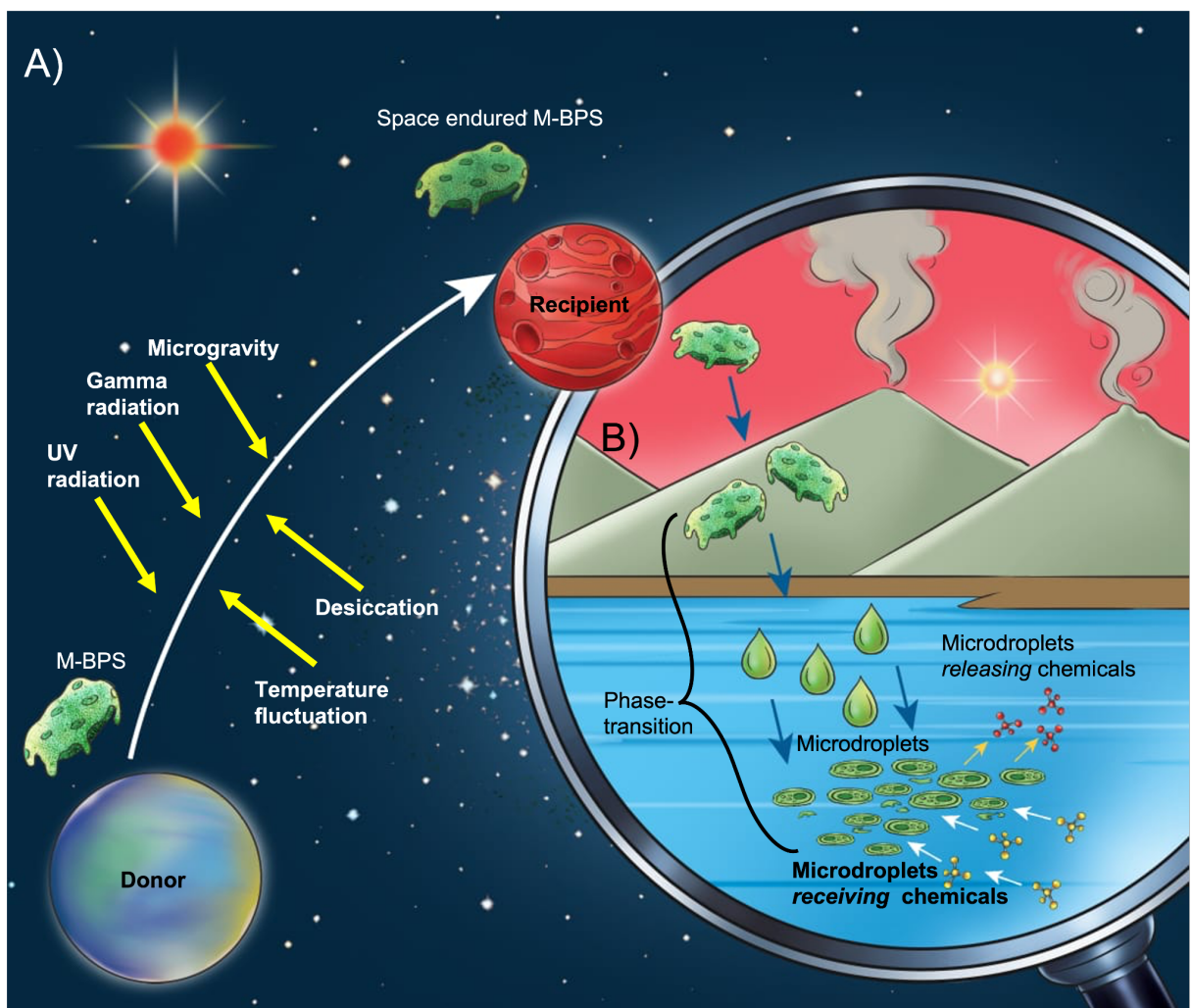

**Figure 1:** Simplified diagram of the Panspermia hypothesis with the emphasis on the OoL by using a material-based Panspermia seed (M-BPS). A) a M-BPS that escapes from a particular planet (donating planet) *via* meteorites or other interstellar objects (omitted for clarity). The "seed" is then scattered into space, endures space exposure, and eventually lands on another habitable planet (receiving planet). B) zoomed-in view of the receiving planet, where a M-BPS enters into a body of water (or other liquid) of the receiving planet, and then undergoes a phase transition to phase separate into microdroplets, which have been proposed as primitive compartments. These microdroplets can either release chemicals to their new ambient surroundings (either through degradation and diffusion of the chemicals it is composed of, or transport of any contents it had already encapsulated beforehand on the donor planet) or receive ambient (local) chemicals from its surroundings by segregating and compartmentalizing. Either of these processes ensures that M-BPS microdroplets have blended into their new environment, enabling chemical continuity towards developing nascent biology, assuming that the microdroplets themselves are stable to degradation during the transport process and in the new environment.

## 3. A potential material-based Panspermia model

M-BPS can be composed of many types of potential polymers, ceramics, and composites, but we will focus on the usage of polymeric (or biopolymeric) gels due to their plausible prebiotic relevance and their particular usefulness in the OoL[109–111].

One way that a polymeric gel material could have promoted the emergence of early life would have been through assembly into primitive compartments, encapsulating primitive metabolic, genetic and catalytic materials on the donor planet[112,113]. Such compartments providing encapsulation functionality are absolutely essential to the emergence and evolution of early life, as they provide a number of essential functions to a primitive system such as material exchange, recombination, and evolution[114,115]. However, primitive compartments come in many sizes and shapes, including lipid bilayer vesicles, membraneless droplets, and even mineral pores, with each providing their own unique advantages and disadvantages. For example, lipid bilayer vesicles can provide stable compartmentalization to encapsulated genetic polymers[116,117], which would have promoted genetic evolution. However, depending on the composition, some lipid compartments are unstable in high salinity[118] or extreme pH[119,120], although this instability can be ameliorated through introduction of divalent cation chelators (like citric acid)[121] or increasing membrane diversity[122–124]. Membraneless droplets generated by liquid-liquid phase separation (LLPS)[115,125], such as coacervates[126] or aqueous two-phase systems[127], have also shown the ability to segregate primitive biomolecules such as RNA or peptides[128–131]. While such systems can be cyclically assembled and disassembled (for example through modulation of environmental conditions such as pH, salt, or temperature[132–134]), depending on the composition, membraneless droplets may have been more "leaky" than vesicles[117]. Other primitive compartments, such as mineral pores, are very stable on long timescales and have been shown to promote polymerization of primitive genetic materials such as RNA[135,136], but their structure is governed by its geochemical composition, and thus it is difficult to envision any dynamic structural changes on the short term. Thus, as one can see, while compartmentalization as a principle must be considered in prebiotic systems, the potential structures that can achieve such compartmentalization are still large in variety and there is no clear consensus within the field.

As such, it is necessary to investigate other types of primitive compartment structures. Polymeric gels in particular could have potentially formed compartments similar to the membraneless droplets described above on prebiotic Earth, potentially through phase separation (or more appropriately, liquid-liquid phase separation (LLPS)), which is a process in which liquid mixtures spontaneously separate into two liquid phases[137], forming membraneless droplets. For example, polyester gels generated from dehydration synthesis (by heating) of alpha-hydroxy acid monomers have recently been shown to form membraneless microdroplets upon hydration in aqueous solution[128,129,138] (Figure 2). Other primitive polymers formed through dehydration polymerization have also shown similar behavior in their ability to form membraneless droplets[139]. Such a mechanism leading towards droplet assembly could have potentially occurred in primitive evaporative environments on early Earth with mild temperature fluctuations, such as hot spring environments[140], as a result of wet/dry cycles due to day/night[141] or seasonal cycles[142], or in the presence of deliquescent materials[143]. Emergence of the ability for membraneless droplets to have been generated from such polymeric gel materials (which simply requires an aqueous medium), both on a donor planet and a receiving planet, could have resulted in a number of functions that promoted further chemical evolution being imbued into a primitive chemical system such as material exchange (as a mechanism for nutrient input, waste output), motion (potentially due to diffusion), coalescence (as a mechanism for recombination), as well as scaffolding amphiphile assembly on droplet surfaces (*i.e.*, evolution towards a membrane-bound compartment state). We also note that while the phase separation that gels undergo to form droplets have so far been shown in aqueous environments, other liquids (such as organic solvents, some of which are abundantly present in other extraterrestrial environments such as the methane lakes of Titan[144]), could also result in a similar phase transition and would depend on the chemistry of the polymeric gels themselves.

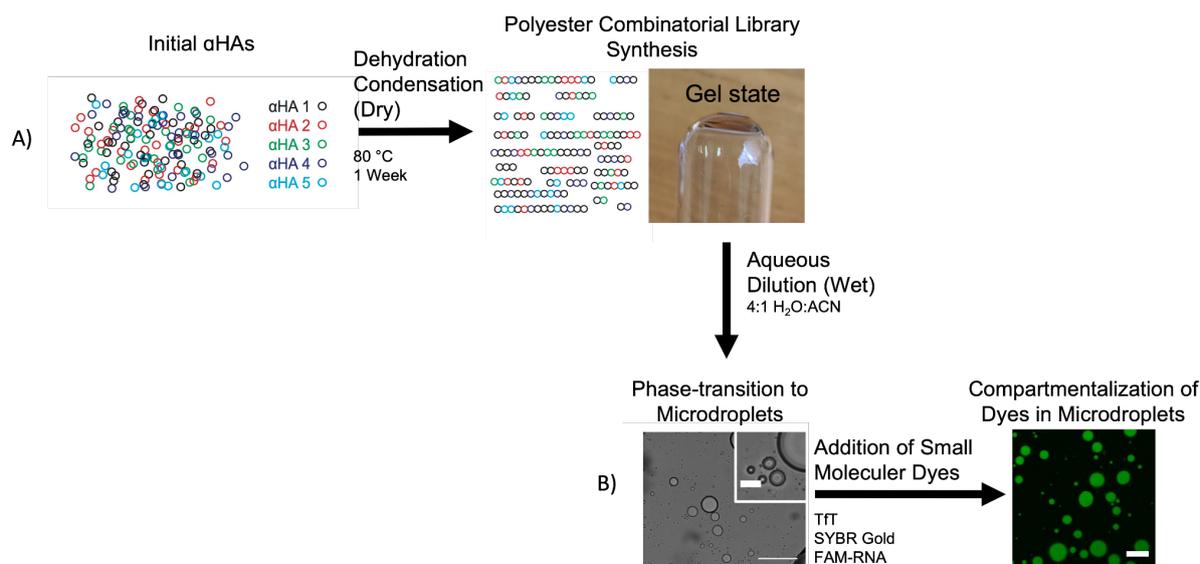

Figure 2: A shows the formation of a dynamic combinatorial library of polyesters using several types of alpha hydroxy acids (αHA) as starting monomers by dehydration [145], and B shows their subsequent microdroplets formation when diluted in an aqueous solution (4:1, H$_2$O: Acetonitrile (ACN)) and its capacity to compartmentalize dyes (Tft, SYBR Gold = generic fluorescent dyes, FAM-RNA = fluorescent-labeled RNA)[128]. This figure is modified from with an exclusive License to Publish to National Academy of Sciences[128] (Jia et al, 2019) (Copyright Jia, T.Z., Chandru, K., et al.), respectively.

### 3.1 Material Exchange

Membraneless droplets formed after phase separation of polymeric gel materials could exhibit a variety of dynamic behaviors that could be indicative or relevant to life. One important aspect is material exchange, as the dynamic and reversible physical property of membraneless droplets shows that encapsulated molecules inside have the ability to dynamically exchange its material with the surrounding environment[146,147]. At the origins of life, such membraneless droplets have been proposed as primitive compartment systems in which internal and external components exchange across the membraneless boundary with lower energetic cost compared with a membrane[131]. In particular, when the external components have chemical affinity and compatibility (charge, polarity, etc.) for the interior of the LLPS, the material exchange process could lead to the spontaneous multiple orders-of-magnitude increase in concentration of DNA[133,148] or of magnesium ions, nucleotides, and RNA within the droplet[130]. Membraneless compartmentalization based on polymers produced by prebiotically available organic compounds such as α-hydroxy acids showed preferential and differential segregation of small molecules and RNA, providing a potential strategy to facilitate primitive chemistries through exchange and encapsulation of primitive components[128]. For example, a M-BPS in droplet form could have uptaken chemicals on an aqueous donor planet, before transitioning to a gel-based form (following dehydration) and transport to a receiving planet (assuming that such encapsulated chemicals could have survived the journey). Then, upon arriving at the receiving planet and phase separation into the droplet form (depending on the environmental conditions at the time), the pre-encapsulated chemicals could then be released into the receiving planet environment.

Finally, similar to the material exchange process of membraneless droplets, biomolecular import/export is significant for facilitating dynamic cell-cell communication in extant biology[149].

Accordingly, communication between membraneless droplets through material exchange can provide transport of genetic/catalytic materials to adjacent compartments, resulting in exchange of chemicals essential for survival in a harsh environment on early Earth.

*3.2    Motion*

Another important aspect of polymeric gel material droplets is motion, which would allow the droplet to move to other locales on the recipient planet which might be more favorable. For example, a different location could result in less degradation of the droplets themselves, while there may be more chemical abundance at a different location for uptake and compartmentalization. Finally, combining this motion with their material exchange properties would allow local transport of chemicals in the recipient planets' environment, potentially facilitating combinations of chemicals and subsequent reactions essential for the origins of life (that would not have been possible beforehand without the M-BPS).

For instance, the surface motion behavior of LLPS droplets has been investigated on a transparent glass surface through fluorescence microscopy[150]. Two surface diffusion modes exist (fix mode and diffusion mode) and the distribution of these two diffusion modes can be tuned by chemical modification of the glass surface with positive charges due to Coulomb interactions between such droplets and the solid surface. The droplets studied contained a negative ζ-potential and consisted of peptides and RNA, but it is plausible for gel-material-based membraneless droplets to exhibit the same properties and, depending on their ζ-potential (which can be tuned based on their composition) encapsulate different types of molecules.

Meanwhile, the distribution of charges in LLPS droplet components also affects droplets assembly (in this case, also made up of peptides and RNA) and their diffusion ability; such a process is related to motor neuronal cell death in Amyotrophic Lateral Sclerosis (ALS) disease[151]. According to Fluorescence Recovery After Photobleaching (FRAP) measurements reflecting diffusion efficiency, droplets of peptide variants with small charge periodicity have a faster FRAP recovery rate than that with larger periodicity, suggesting differences in internal exchange. As such, due such differences in physical diffusion processes, component molecules can be rearranged within the droplets. While again, such a demonstration is based on non-gel-material-based droplets, previous demonstrations showed the ability for primitive gel-based droplets to acquire differential charges based on residue incorporation.[128,129]

*3.3    Coalescence*

For modern cells experiencing "cell cycles", the coalescence process is required to undergo repair, reproduce, and distribute their genetic information to prevent genetic extinction[152]. In principle, membraneless droplets adopt round morphologies with minimal surface tension, and coalesce into a single droplet upon contact with each other[153]. There are mainly two mechanisms of droplet coalescence process. Ostwald ripening results in small molecules moving to "find" larger droplets due to the minimization of surface free energy, resulting in molecules located close to the surface of droplets energetically less stable to those in the droplet interior[154]. On the other hand, passive coalescence occurs when droplets move freely *via* Brownian motion, and when two droplets similar in size contact each other randomly, they then fuse into a single larger droplet[155,156]. Preliminary results show that NaCl concentration increases induce membraneless polyester droplet coalescence, potentially due to migration of ions to the droplet interface, resulting in surface tension or surface charge changes[128,157], while temperature increases can also accelerate membraneless droplet coalescence[158]. The rate of coalescence depends on aqueous solution properties as well, such as phase volumes and viscosities[159]. However, unilamellar lipid vesicles[160] and natural clay microparticles[161] have been put forward as a mechanism to stabilize LLPS droplets as artificial cell model at interface, preventing or controlling droplet coalescence without serving as a barrier to the input/output of molecules up to the size proteins or nucleic acids in some cases[160]. Hence, the initiation/control of polymeric gel material droplets through prebiotic environmental factors could be a significant mechanism leading to their growth or division.

## 3.4 Evolution

Evolution of a primitive compartment system with limited function into a complex compartment system, similar to a cell, may have been one mechanism by which the first cells emerged on early Earth[115]. However, this is not the only mode of primitive compartment evolution, as it may be possible for primitive compartments to evolve more function even if they do not ultimately become more "cell-like"[138]. While it is plausible for a polymer gel material compartment to have evolved through sequence replication and evolution of an encapsulated genetic polymer (such as a nucleic acid)[162], it is also possible for such compartments to evolve in other ways. As such, here, we briefly discuss two modes by which polymer gel material droplets could have evolved: compositionally through evolution of its polymer sequence or structurally through acquisition of more complex structural components.

Compositional evolution of polymeric materials, which has been proposed to have been plausible before the advent of genetic biopolymers[163,164], would have required that the composition of any material directly affects its "phenotype", and that modulations of this composition result in evolution of the material. In particular, this means that the various properties governing polymeric gel assembly, structure, and/or function are directly controlled by its polymer composition, and that modulation of the polymer composition (whether temporarily or permanently) may result in differences in the material properties. For example, recent observations suggested that modulation of the composition of a polyester microdroplet to include more basic residues resulted in acquisition of RNA segregation and intrinsic fluorescence[129]. While this is just one observation, the potential exists for additional compositional modulations of polyester microdroplets to affect the compartment assembly, structure, or function in other unknown ways. As such, perhaps through cyclical polymerization/hydrolysis events facilitated by wet-dry cycles[129,138,165], one may be able to experimentally demonstrate evolution of polymeric gel-based material compartments.

Structural evolution of compartments may have resulted in acquisition of more complex structural components which contributed more complex functions to the compartment system, such as demonstrated *via* coacervate acquisition of liquid crystal character[148,166] or lipid membranes[167–169]. Emergence of such structures in polymeric gel material droplets could have occurred, perhaps through acquisition of the nucleic acids or lipids that formed those structures. In particular, such compartments would have benefited from the ability to actively scaffold the assembly of additional structures (rather than passively acquiring such structures). For example, hyperbranched polymers have been shown to be able to scaffold the assembly of zinc sulfide nanocrystals[170]. For example, perhaps polymeric gel compartments could have actively scaffolded lipid assembly on its surface[128] to promote the formation of an enclosing membrane. This ability for a membraneless droplet to actively acquire and assemble a lipid membrane could have been a key stage of evolution connecting a membraneless protocellular world (polymeric gel material droplets) to a membrane-bound world (more cell-like state)[126].

We described here plausible mechanisms by which polymeric gels can potentially be applied as a M-BPS, especially through supramolecular structures generated from LLPS, that can potentially drive chemical evolution upon arrival on a recipient planet. However, we must bear in mind that such proposed mechanisms, specifically on how chemical evolution would occur on an extraterrestrial planet, are vague considering that researchers still understand very little about how chemical evolution led to terrestrial OoL in general. The problem of understanding terrestrial OoL is multifold (discoursed elsewhere, *e.g.*,[1,171–174]), and, a major oversight in all OoL models (*e.g.*, RNA world, metabolism first, *etc*), although compatible with the structures and functions of extant life[175], is that OoL models generally appear to minimize the role of historical contingencies[74,176], *i.e.*, open-ended circumstances which undoubtedly shaped the direction or plausibility of the OoL, but cannot be predicted *a priori*. For instance, contingencies that shaped the OoL on Earth may have come, among many, in the form of certain emergent properties of prebiotic chemistry driven by varying geochemical processes (*e.g.*,[177]), or global/local geological changes[178] that can potentially change inanimate matter to animated matter, *i.e.*, life-as-we-know-it. In fact, contingencies were (and still) are instrumental after the onset of Darwinian evolution[179] (*e.g.*,[180,181]). Hence, in the same vein, such contingencies will also shape the direction of how a M-BPS drives chemical evolution on a new planetary environment (if at all). While there are a number of remaining open questions pertaining to material-based Panspermia models, how to account for contingencies in chemical evolution is by far the most difficult open question that remains to be explored experimentally.

## 4. Open Questions and Limitation to the material-based Panspermia hypothesis

Overall, the ability of polymeric gel materials to form droplets in an aqueous conditions *via* LLPS to possess certain functionalities that could have contributed to the emergence or evolution of life (material exchange, motion, coalescence, evolution, *etc.*) makes them desirable candidates as M-BPS. Given that early planetary systems were turbulent, where land mass from a certain planet can be transferred to another *via* meteoritic transfer (*e.g.*, during late heavy bombardment (LHB)[182]), polymeric gels could have plausibly been transported between nearby planets (*e.g.*, Mars and Earth, the outer moons of Jupiter and Saturn[183,184] and the TRAPPIST-1 system[3]). However, although a polymeric gel material-based M-PBS is conceptually palpable, there are several open questions and limitations about the plausibility of this concept that deserve considerable attention before its full application.

A major question is: can M-BPS survive the harsh conditions in space during transport? For example, prebiotically plausible polyester gel materials, similar the DNA mentioned above, may succumb to space degradation (*e.g.*, by microgravity, UV radiation, temperature, cosmic rays, *etc.*) resulting in their degradation or hydrolysis, and hence determining its rate of degradation in such conditions is warranted. These experiments can be done in a ground-based facility simulating a space environment (*e.g.*,[185,186]) and/or exposing the samples to LEO on the International Space Station (*e.g.*,[16,38,41]). In addition, other aspects related to their degradation must also be investigated. For instance, what are the limits of degradation that these polymeric gel materials can tolerate and still retain their functions? Second, can the degradation rates of the polymeric gel materials be minimized if they are combined with inorganic or meteoritic matter (*e.g.*,[52,187]) as a proxy of lithopanspermia? Previous studies have shown that some organisms and organic chemicals undergo little degradation when associated with minerals or meteoritic matter. For example, *Bacillus subtilis*'s spores survived better (up to 5 orders of magnitude) in spaceflight exposure of 2-weeks when mixed with powdered clay, rocks and meteorites[44]. And in a separate study, trimer of leucine and diketopiperazine (DKP) (a cyclic dimer of glycine), when associated with powdered meteorite film (Allende), also degrade lesser when exposed to spaceflight for 98 days on board the MIR space station[104].

Beyond degradation during spaceflight, meteorite impact shock (*e.g.*, ejecta impacts on Mars, the Moon and Earth during the LHB period[188]) may also degrade polymeric gel materials (such as polyester gels) upon their arrival onto a recipient planet, which requires investigations on chemical and mechanical degradation of such materials caused by impact shocks. However, given that organisms such as *Tardigrada*[189] and some bacteria (*e.g.*, *Bacillus subtilis* spores and *Deinococcus radiodurans* cells)[190] have been experimentally shown to survive low to moderate impact shocks, it is possible that polymeric gel materials have the potential to survive similar or even greater-strength shocks during transport.

Another bottleneck is that phase-separation of polymeric gel materials (polyesters) from previous works (mentioned in section 3.3) were investigated only in aqueous conditions mimicking early Earth environments, either with pure $H_2O$[129] or $H_2O$ together with a small ratio of organic solvent (*e.g.*, Acetonitrile)[128,139]. However, when it comes to "seeding" a new planet using polymeric gels as M-BPS, $H_2O$ may not be the only solvent present (*e.g.*, Titan's hydrocarbon lakes[191]). Hence, testing the gels' ability to phase-separate with different solvents, both aqueous-based (including with a wide range of pH and salinity) as well as organic-based, is necessary. However, it has been found that other planetary bodies, mainly Europa and Enceledus, have shown strong evidence of water[192] below the surface, *i.e.*, subsurface oceans, suggesting that as long as the polymeric gels can reach such aqueous environments, that they have the potential to phase separate. However, it is also plausible that the cold temperatures encountered during space travel and on other planetary surfaces may result in irreversible structural damage, and further investigations into the robustness of the structure polymeric gel materials at extreme temperatures must be undertaken.

Finally, one of the most pointed criticisms towards experimental validation of Panspermia hypotheses is that the experiments conducted only encompass short time scales which do not represent realistic astronomical (or geological) timescales. For instance, the longest Panspermia-related experiment, to our knowledge, was performed over 6 years[13]. Such experimental durations fall short of the time needed for Panspermia seeds to be transferred between planets. For example, the travel time of meteorites between Mars and Earth is ~ 600,000 to 14 million years[86], and thus it is, at least in the present, humanly impossible to perform experiments matching such time scales. This problem is in fact synonymous with OoL research in general[173], where the OoL was estimated to have occurred over 500,000 - 1 billion years[193–195], yet the laboratory proxies, *i.e.*, prebiotic chemistry experiments, are often (but not always) done in much shorter timescales (*e.g.*, hours to days to weeks, *etc.*). Because of this time-scale issue, the role of making assumptions and speculations on the OoL based on such short-term experimental simulations is rather unavoidable even though they are firmly grounded (and updated) on what we presently know about biology and geochemical constraints of early Earth and other planetary bodies. Nevertheless, efforts are being made to address this problem through longer-term prebiotic chemistry investigations (*e.g.*,[196]).

## 5. Conclusion

To our knowledge, experimental validation of Panspermia hypotheses performed thus far overtly tests the limits of life in extreme conditions, *i.e.*, spaceflight, but stops short of explaining the "seeding" part of the hypothesis. While these studies are interesting and full of prospects to study the coping mechanism of Earth-bound biology in extreme conditions, it is uncertain that such biological seeds can lead to life on a new planet.

Instead, we propose an alternative material-based Panspermia hypothesis utilizing M-BPS in the form of polymeric gel materials that can potentially be an experimental framework to study the Panspermia hypothesis and the OoL. We have briefly shown that polymer gels (in the form of polyesters) formed from a variety of prebiotically available hydroxy acids[129,145], amino acids and cyclic compounds[139] showed cell-like functionality and have the potential to lead to the emergence of life *via* chemical evolution. While these polymeric gels' cell-like functions[128,129,139] provide glimpses of the potential of non-biomolecule-based OoL models (*e.g.*,[138,175,197]), its robustness as a Panspermia seed, especially during spaceflight and also impact shock upon arriving in a recipient planet is unknown, requiring necessary investigation. As such, with further experimental study and validation, the plausibility of M-BPS and polymeric material-based Panspermia hypotheses can be utilized as a prototype to explore possible means to terraform other planets and glean more information about aspects of the OoL not yet investigated.

**Competing Financial Interests:** None declared

**Acknowledgments:**

We would like to thank the following people for their initial discussions in relation to this manuscript: Dr. Chaitanya Giri (ELSI), Professor Jalifah Latiff (UKM), Dr. Wee Boon Hi (UKM), Professor. P. Susthitha Menon (UKM), Dr. Daniel Law Jia Xian (UKM), Dr. Ahmad Daniel (University of Malaya), Professor Terry Kee (University of Leeds), Professor Kensei Kobayashi (Yokohama National University), Professor Hajime Mita (Fukuoka Institute of Technology), and Dr. Affifudin Hussair bin Mat Jusoh (UKM). We would also like to thank the organizers of the American Chemical Society spring meeting Polymer Colloids Symposium for the opportunity to connect with the journal editor for this opportunity.

This work was supported by the Malaysian Ministry of Education (FRGS/1/2021/STG04/UKM/02/1).

C.C. and T.Z.J. are members of the Earth-Life Science Institute (ELSI) of Tokyo Institute of Technology, which is sponsored by a grant from the Japan Ministry of Education, Culture, Sports, Science and Technology (MEXT) as part of the World Premier International Research Center Initiative (WPI). T.Z.J. is supported by Japan Society for the Promotion of Science (JSPS) Grants-in-aid 18K14354 and 21K14746, the Tokyo Institute of Technology Yoshinori